\documentclass[final,11pt,3p,sort&compress]{elsarticle} 
\makeatletter
\def\ps@pprintTitle{%
 \let\@oddhead\@empty
 \let\@evenhead\@empty
 \def\@oddfoot{\reset@font\hfil\thepage\hfil}%
 \let\@evenfoot\@oddfoot}

\biboptions{super}

\usepackage[T1]{fontenc} 
\usepackage[latin2]{inputenc}
\usepackage{amssymb}
\usepackage{amsthm}

\usepackage{graphicx,cancel}
\usepackage{ulem}
\usepackage{color}
\usepackage{fixltx2e} 
\usepackage{dblfloatfix} 
\usepackage{booktabs} 
\usepackage{microtype} 
\usepackage{hyperref}
\hypersetup{
    colorlinks=true, 
    linkcolor=blue,  
    citecolor=blue,  
}
\usepackage[all]{hypcap}  

\renewcommand{\_}[1]{{}_{\mathrm{#1}}}

\begin{document}

\begin{frontmatter}

\title{Manipulation of the electronic structure by reversible dehydrogenation of tetra(\textit{p}-hydroxyphenyl)porphyrin molecules}

\author[afko]{Lars~Smykalla}
\author[afko]{Pavel~Shukrynau}
\address[afko]{Technische Universität Chemnitz, Institute of Physics, Solid Surfaces Analysis Group, D-09107 Chemnitz, Germany}
\author[chem]{Carola~Mende}
\author[chem]{Tobias~R\"uffer}
\author[chem]{Heinrich~Lang}
\address[chem]{Technische Universität Chemnitz, Institute of Chemistry, Inorganic Chemistry, D-09107 Chemnitz, Germany}
\author[afko]{Michael~Hietschold}

\date{\today}

\begin{abstract}

The controlled and reversible interconversion between the free-base and the doubly dehydrogenated form of a 5,10,15,20-tetra\-(\textit{p}-hydroxy\-phenyl)\-porphyrin molecule in an ordered array is demonstrated. This is achieved through voltage pulses by hydrogen transfer between the center of the porphyrin and the tip of a scanning tunneling microscope (STM). The local dehydrogenation leads to significant shifts in the energetic positions of the molecular orbitals. Density functional theory (DFT) calculations  corroborate our conclusions and allow to gain more insight into the different energy level alignment before and after dehydrogenation. Due to the different conductance at a given voltage a clear distinction of both molecular species is possible, which also enables the application as a single-molecular switch.

\end{abstract}

\end{frontmatter}

\section{Introduction}

Molecular-based nano-electronics is a field of rapidly growing interest. The well-defined preparation of organic molecules at surfaces and in nano-scale environments is a prerequisite to exploit their molecular properties both in single-molecule devices and highly organized supramolecular architectures.\cite{Yoshimoto2010, Stepanow2008, Hietschold2005}
Perhaps one of the most basic molecular devices is a binary one. Molecules with an on/off functionality, e.g. two states of lower and higher conductivity, have potential applications in, for example, nano-scale memory devices or molecular logic gates.\cite{Green2007}
For such devices, it is required that two stable states can be reversibly switched in a controllable manner. This is done in many cases by inducing conformational changes in a molecular system \textit{via} external stimuli such as light\cite{Comstock2007}, short-range force, an electric field\cite{Gopakumar2006b} or electron injection, e.g. from a STM tip\cite{Moresco2001, Qiu2004, Iancu2006, Komeda2011, Pan2009, Fu2009, Liljeroth2007, Auwaerter2012}.
Another route to achieve a molecular ``switch'' is in the sense of interconversion between different electronic or magnetic properties by controlled local breaking of chemical bonds \textit{via} a STM tip, especially dehydrogenation reactions.\cite{Katano2007, Li2010, Liu2013} The problem here usually lies in the reversibility as exposure to hydrogen is necessary to restore the initial molecules and their properties.

For the application in devices, it is desirable to arrange the functional molecular units in ordered arrays where the single molecules can be individually addressed and manipulated. From this point of view, porphyrin-based molecules are very promising. They have been studied in great detail for many years because of their high chemical and thermal stability, multiple stable oxidation states~\cite{Muellegger2011a} and exceptional versatility.\cite{Gottfried2009}
In addition, these organic molecules can be easily modified to change their physical properties and mutual interaction.\cite{Ishtaiwi, Yokoyama2004, Fendt2009, Lu2009}

It was shown previously that a local dehydrogenation in the center of tetraphenylporphyrin molecules could be achieved with a STM tip.\cite{Auwaerter2012} Within this study, it is shown how dehydrogenation influences the electronic properties of the molecule and, furthermore, that it is also possible to reverse this reaction. The reversible, and controllable interconversion of individual porphyrin molecules between the free-base and the dehydrogenated form is enabled by using only STM. The manipulation of 5,10,15,20-tetra(\textit{p}-hydroxyphenyl)porphyrin (H$\_2$THPP) molecules in a self-assembled ordered array and with this a switch in the electron transport properties of individual molecules is demonstrated in the following.

\section{Experimental and computational details}

\begin{figure}[tb]
\centering
\includegraphics[width=0.4\textwidth]{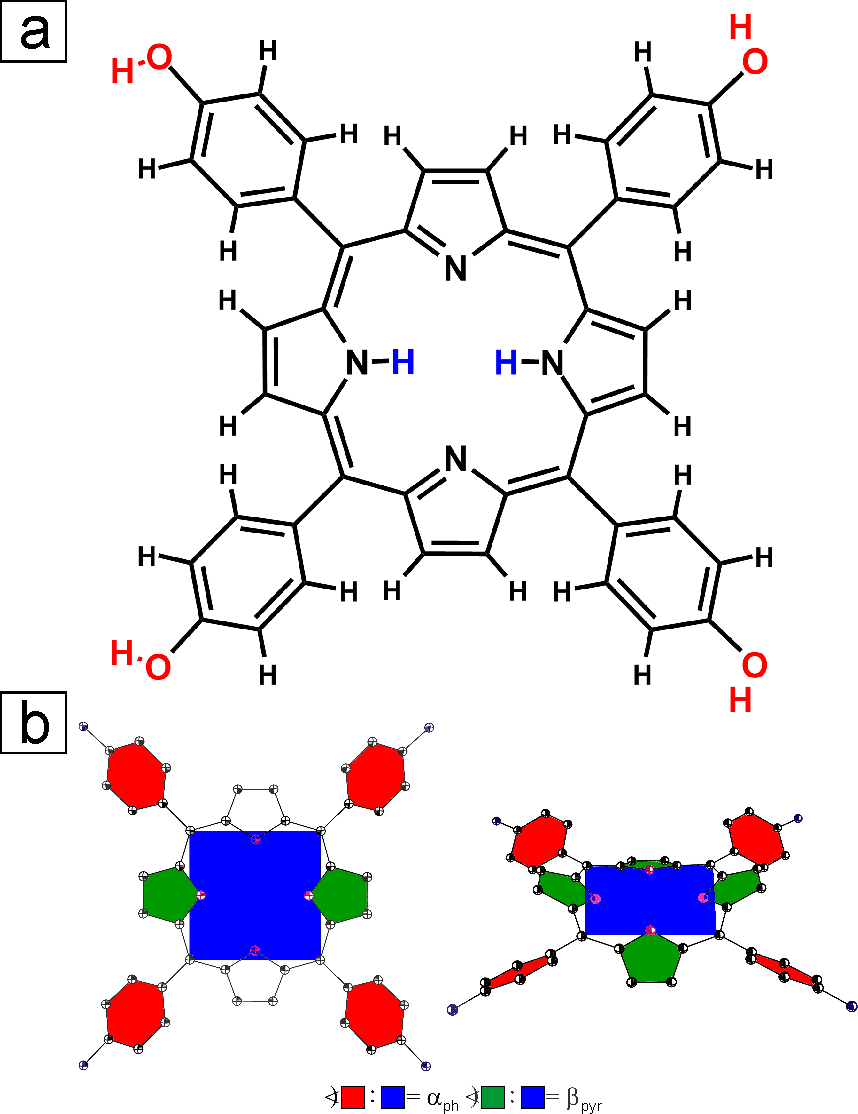}
\caption{(a) Chemical structure of the free-base 5,10,15,20-tetra(\textit{p}-hydroxy\-phenyl)\-porphyrin (H$\_2$THPP) (b) Definition of the angle of rotation of the phenyl ($\alpha\_{ph}$, red) and pyrrol ($\beta\_{pyr}$, green) plane relative to the mean plane of H$\_2$THPP (blue).}
\label{fig:mol}
\end{figure}

H$\_2$THPP was obtained by the direct condensation of \textit{p}-hydroxybenzaldehyde with pyrrole in propionic acid under reflux according to Adler \textit{et al.}\cite{Adler1967} The chemical structure is shown in Fig.~\hyperref[fig:mol]{\ref*{fig:mol}(a)}. A clean surface of an Au(111) single crystal was prepared by multiple cycles of Ar$^+$-sputtering at an energy of 500\,eV and annealing to 400~$^{\circ}$C for 1\,h. H$\_2$THPP molecules were deposited on the substrate by organic molecular beam epitaxy (OMBE) in ultra high vacuum (UHV). For this, solid H$\_2$THPP was first purified by heating to a temperature slightly below the sublimation temperature in UHV for two days. The molecules were then deposited by organic molecular beam epitaxy at around 350~$^{\circ}$C on Au(111) (pressure in the preparation chamber in the range of $10^{-8}$~mbar). The temperature of the substrate during deposition was held at room temperature (ca. 23~$^{\circ}$C). The sample was subsequently annealed to $\approx 150~^{\circ}$C for one hour to promote diffusion of the molecules to achieve an uniform coverage over the whole sample and the formation of large highly ordered areas.
The scanning tunneling microscopy experiments were performed with a variable-temperature STM from Omicron in UHV. The base pressure in the UHV chamber was in the range of $10^{-10}$~mbar. Electrochemically etched tungsten tips for the STM were annealed in UHV by bringing them close to a heated tungsten filament to remove the tungsten oxide layer and remnants of the etching solution from the end on the tip. All measurements were done at room temperature. STM images were measured in the constant current mode with a tunneling current of 100~pA. All voltages refer to the bias with respect to the sample. STM images were processed with the WSxM software~\cite{Horcas2007}, whereby moderate low pass filtering was applied to reduce noise. Density Functional Theory calculations were performed with Gaussian 03.\cite{Gaussian03} The hybrid DFT exchange-correlation functional B3LYP~\cite{Becke1993} and a 6-311+g(d,p) LCAO basis set were used.

\section{Experimental results}

\begin{figure}[tb]
\centering
\includegraphics[width=0.5\textwidth]{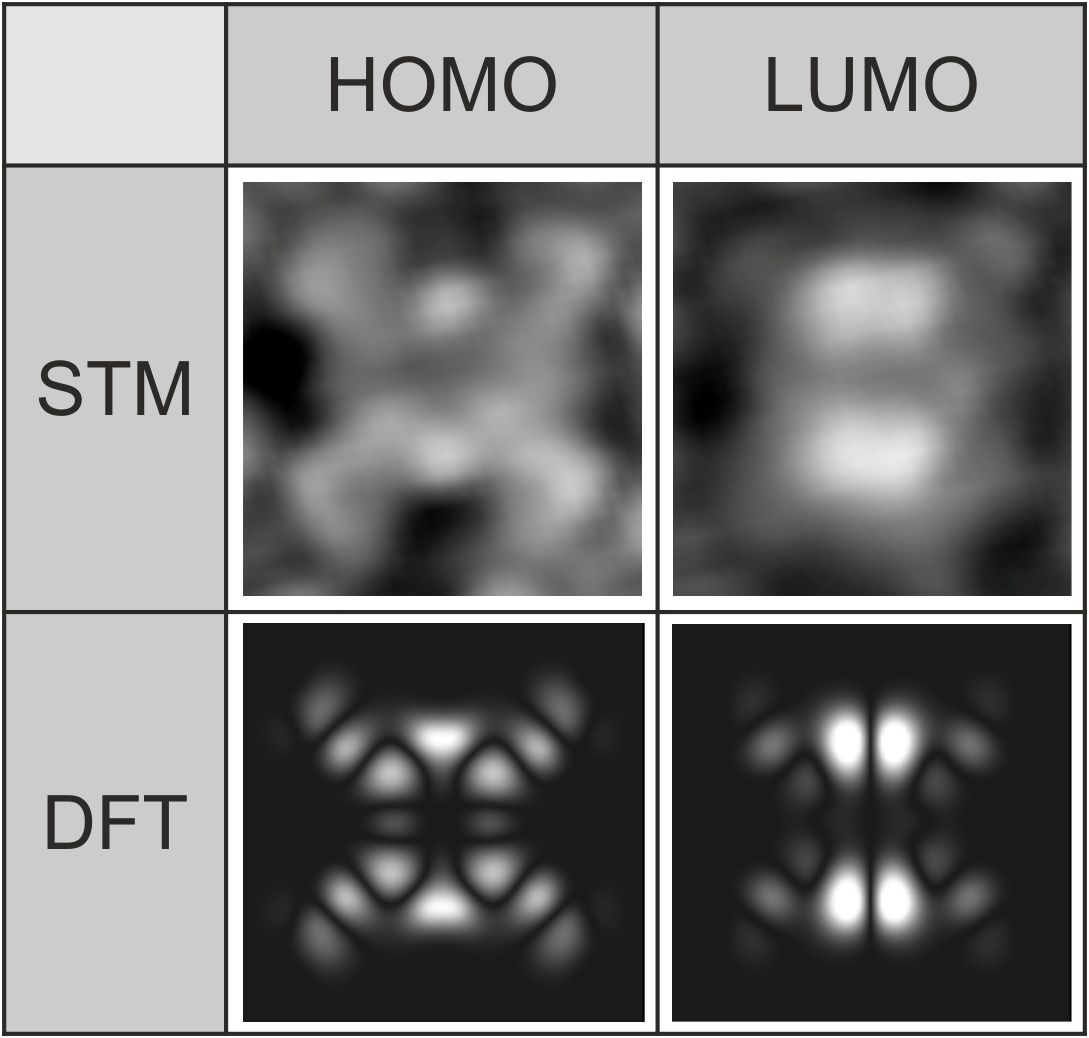}
\caption{Comparison of the measured HOMO and LUMO appearance in STM with the theoretically simulated STM images of saddle-shape deformed H$\_2$THPP ($\alpha\_{ph}=32.5^{\circ}$).}
\label{fig:stm-dft}
\end{figure}

The ordered arrangements of H$\_2$THPP on the surfaces of Au(111) and Ag(110) was reported previously.\cite{THPPstruc} When adsorbed on the surface, H$\_2$THPP shows a saddle-shape deformation of the porphyrin macro\-cycle, as observed in STM (Fig.~\ref{fig:stm-dft}). Such a deformation is usually found upon the adsorption of metal-free and metallo-tetra\-phenyl\-porphyrins (TPP) on metal surfaces.\cite{Auwaerter2012, Ecija2008, Santo2011, Diller2012} Thereby, the phenyl rings are turned toward a more parallel adsorption geometry (small $\alpha\_{ph}$) to increase the interaction of the $\pi$ electrons with the substrate surface by decreasing the distance between the surface plane and the mean porphyrin plane. Through steric repulsion the pyrrole rings of the macrocycle are consequently tilted out of the mean plane ($\beta\_{pyr}$ in Fig.~\hyperref[fig:mol]{\ref*{fig:mol}(b)}) leading to the typical protrusions from C $\pi$ electrons of opposite pyrrole rings as can be seen in Fig.~\ref{fig:stm-dft}. In our highly resolved STM images, the appearances of the highest occupied (HOMO) and lowest unoccupied molecular orbital (LUMO) are in very good agreement with the local density of states which was calculated with DFT (Fig.~\ref{fig:stm-dft} bottom). Also, from the two pairs of protrusions in the LUMO, the azimuthal orientation of the molecular saddle-shape can be easily determined.

\begin{figure*}[tb]
\centering
\includegraphics[width=0.9\textwidth]{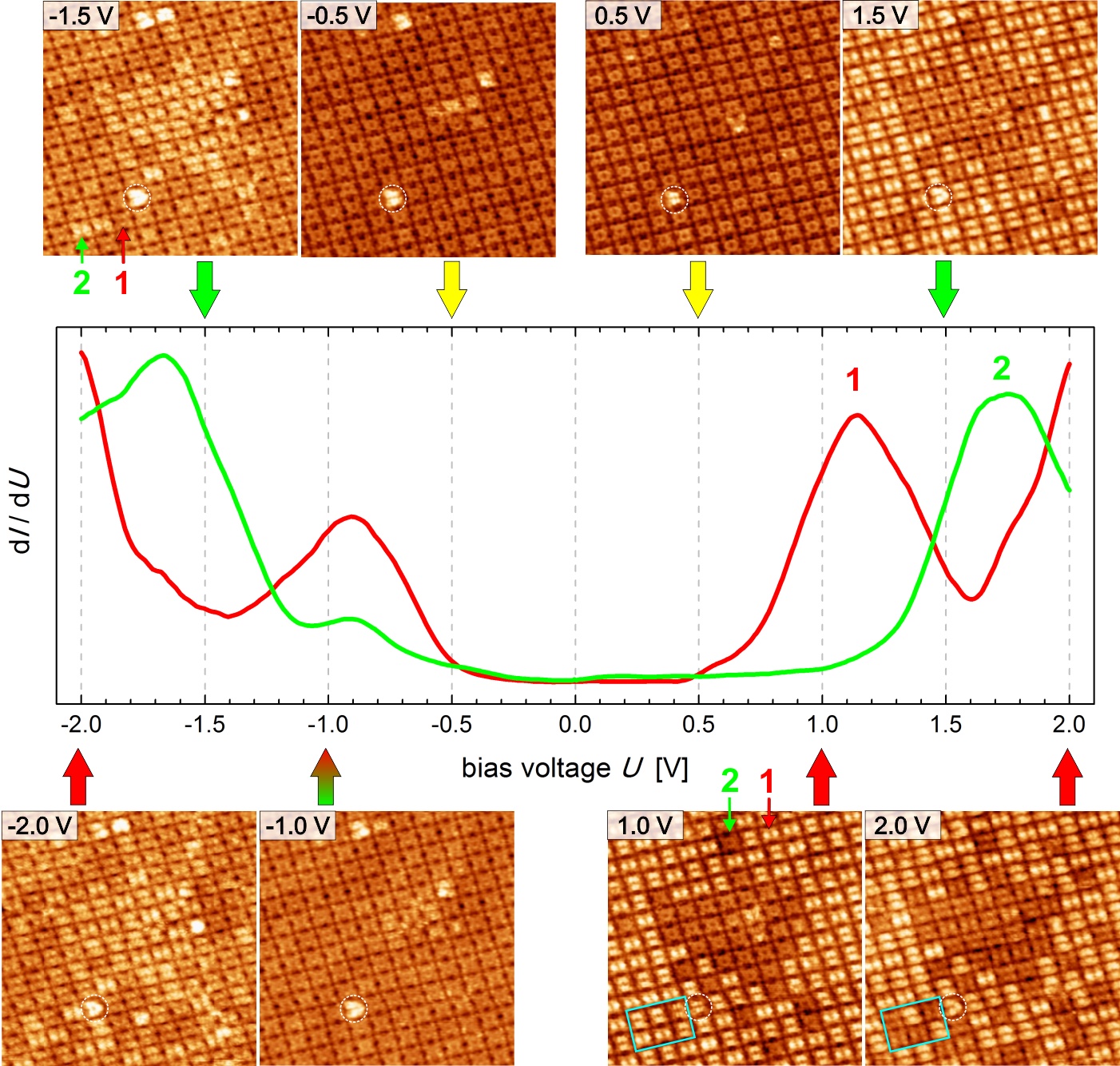}
\caption{Electronic structure: $(\mathbf{d}I/\mathbf{d}U)$ tunneling spectra are shown for each variant \textbf{1} (red) and \textbf{2} (green) of the porphyrin molecule. STM images with the corresponding molecular appearances are shown for the specified voltages (each image is 21\,nm $\times$ 21\,nm). Red/green arrows indicate the correspondence to molecular orbitals of \textbf{1}/\textbf{2}, respectively, yellow arrows stand for in-gap contrast. A single molecule adsorbed on the monolayer is marked with a white circle. An area where the molecular appearance changed during scanning is indicated by the cyan rectangle.}
\label{fig:spannungsbilder}
\end{figure*}

The voltage dependence of the appearances of the molecules in STM was probed (Fig.~\ref{fig:spannungsbilder}), which means the energetic positions of the molecular orbitals relative to the Fermi-energy and the corresponding spatial distributions of the electron density were measured. Two different variants of the deposited molecule, in the following named \textbf{1} and \textbf{2}, were found according to their appearance in the STM images (marked by arrows in Fig.~\ref{fig:spannungsbilder}). Scanning tunneling spectroscopy (STS) was measured for each species. The differential conductance of \textbf{1} (red STS curve in Fig.~\ref{fig:spannungsbilder}) shows a peak at $U=-0.9~\mathrm{V}$ which corresponds to the HOMO and HOMO-1. The next energetically lower lying occupied orbitals appear at around $U=-2.0~\mathrm{V}$. The degenerated LUMO and LUMO+1 of \textbf{1} are located at circa $U=1.15~\mathrm{V}$ above the Fermi energy and the LUMO+2 at circa $U=2~\mathrm{V}$. If we look at $(\mathbf{d}I/\mathbf{d}U)$ of \textbf{2} (green STS curve in Fig.~\ref{fig:spannungsbilder}), it can be seen that a peak corresponding to the HOMO is still measured at $U=-0.9~\mathrm{V}$ but a large peak from the convolution of the HOMO-1 and lower lying orbitals appears with the maximum at around $U=-1.7~\mathrm{V}$. A difference is also observed for the LUMO/LUMO+1 peak which is shifted to higher energy. All fitted energetic positions of the molecular orbitals are also summarized in Table~2 of the supporting information (SI). The HOMO $-$ LUMO gap measured from the centers of the peaks is $(2.0\pm 0.1)~\mathrm{eV}$ for \textbf{1} and $(2.5\pm 0.1)~\mathrm{eV}$ for \textbf{2}. The conductance gap measured from the peak onsets is accordingly circa 0.8~eV smaller.
Due to the different electronic structure, \textbf{1} and \textbf{2} can be distinguished by their appearances in STM depending on the applied bias voltage, as shown in Fig.~\ref{fig:spannungsbilder}. For bias voltages inside the HOMO $-$ LUMO gap (e.g. $-0.5~\mathrm{V}$ and $+0.5~\mathrm{V}$), where geometry effects dominate the molecular appearance in STM, all molecules in the ordered layer look exactly identical. At these voltages, single molecules adsorbed on top of the first layer can be discerned due to their larger height (white circle, see also Fig.~S2 in the SI). The LUMO shift is clearly revealed in the STM image: If the molecular layer is scanned at $+1.0~\mathrm{V}$, the previously discussed large protrusions from $\pi$ electrons of the LUMO become visible only for \textbf{1}, whereas the appearance of \textbf{2} is unchanged. For the STM image at +1.5\,V, tunneling also into the LUMO of \textbf{2} becomes possible and is accordingly visible in the molecular appearance. For the filled states, first at the energy of the HOMO, \textbf{1} and \textbf{2} are indistinguishable. In the STM image measured at $U=-1.5~\mathrm{V}$ tunneling from the HOMO-1 and also energetically lower lying molecular orbitals of \textbf{2} become involved. This leads to an increased apparent height of \textbf{2} compared to \textbf{1} at this voltage. For the scan at $-2.0~\mathrm{V}$, this height difference decreases again because additional occupied orbitals of \textbf{1} at this voltage contribute to the tunneling current.
From the molecular appearances of \textbf{1} and \textbf{2}, it is evident that both species must have the same geometry with a saddle-shaped macrocycle.

While scanning with an applied bias voltage higher than about $1.5~\mathrm{V}$, the sudden conversion of individual molecules from \textbf{1} to \textbf{2} is observed. This can be seen when comparing the marked area in the STM images obtained at $+1~\mathrm{V}$ and $+2~\mathrm{V}$ in Fig.~\ref{fig:spannungsbilder}. With increasing bias voltage an also increasing probability of the tip-induced change of H$\_2$THPP molecules is observed, which also depends on the sharpness of the tip~\cite{Auwaerter2012}. For example for the measurement shown in Fig.~\ref{fig:spannungsbilder}, at $+1.5~\mathrm{V}$ and 100~pA a \textbf{1} $\rightarrow$ \textbf{2} conversion rate of 0.01~s$^{-1}$ and at $+2.0~\mathrm{V}$ and 100~pA of 0.2~s$^{-1}$ could be estimated by considering the scan speed, the size and the number of changed molecules. It should be noted that, while measuring at around $+1~\mathrm{V}$, no conversion could be observed but both species can already be discriminated.
The measurement of previously ``untouched'' areas with this bias indicated that \textbf{2} was already present in the monolayer ($\approx 5\%$), which could be due to annealing of the sample to achieve large-scale supra-molecular ordering.

\begin{figure}[tb]
\centering
\includegraphics[width=0.6\textwidth]{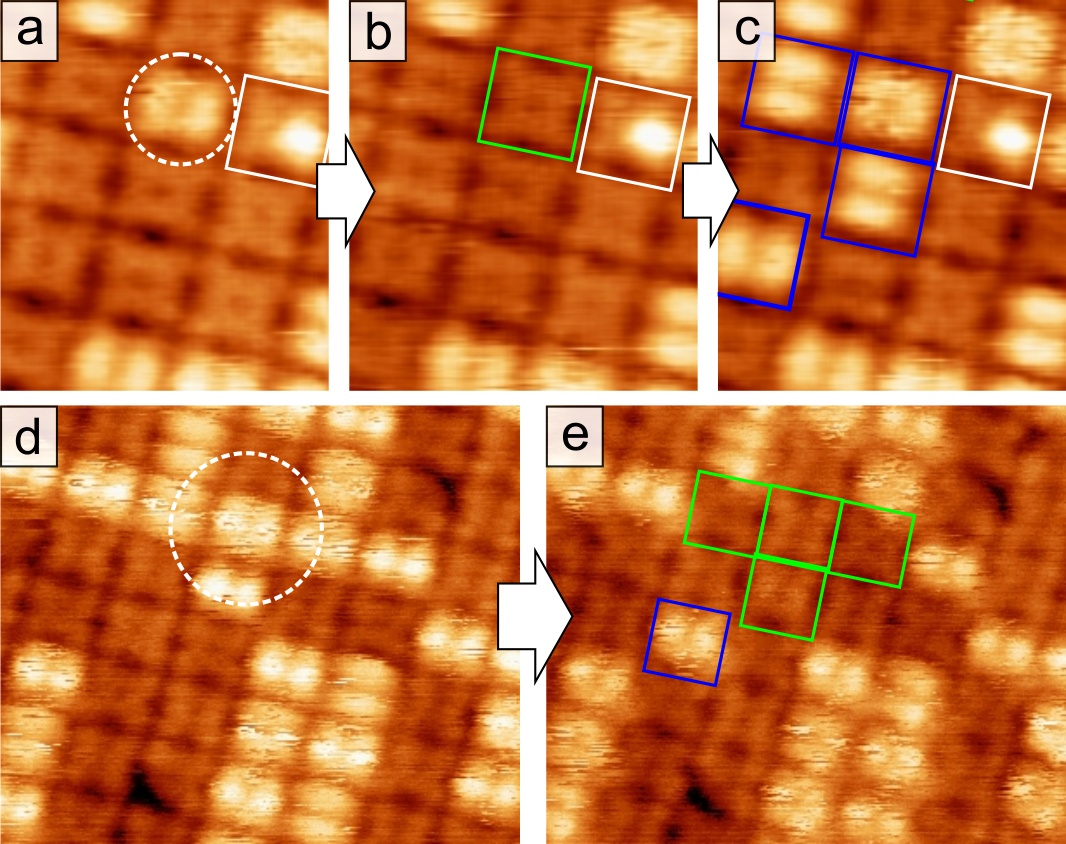}
\caption{Manipulation of the electronic structure by applying a voltage pulse of 2~V for 1~s at the position marked with a white circle. Converted molecules are marked with a green (\textbf{1} $\rightarrow$ \textbf{2}) or blue (\textbf{2} $\rightarrow$ \textbf{1}) rectangle. (a) to (c): \textbf{1} $\rightarrow$ \textbf{2} $\rightarrow$ \textbf{1} conversion with 2~V pulses at the same position before each frame. A molecule with only one protrusion (white rectangle) is used to mark the position in the scanned area. (d) to (e): Changing of neighbouring molecules in both directions with the same pulse (imaging parameter: +1.2~V, 100~pA; top: circa 6~nm $\times$ 7~nm, bottom: 11.5~nm $\times$ 10~nm).}
\label{fig:schalten}
\end{figure}

Inter-conversion events can be induced in a controlled manner by voltage pulses with the tip positioned over the molecule, as demonstrated in Fig.~\ref{fig:schalten}.
This is possible for positive as well as negative bias voltages. Furthermore, the conversion rate strongly increases for a larger set point value of the tunneling current (see the $I(t)$ traces (open feedback loop) in Fig. S1 of the SI)
With a $|U|=2~\mathrm{V}$ pulse a molecule can be changed from \textbf{1} to \textbf{2} with a high converting probability. However, at such a high voltage often adjacent molecules are also converted (bottom images of Fig.~\ref{fig:schalten}). This could be due to an increased effective area of the electron beam between tip and molecular layer or energy transfer from the addressed molecule to its neighbors. Interestingly, changing \textbf{2} to \textbf{1}, that is to brighter appearance at positive bias, is also possible with a voltage pulse of the same polarity. This can occur by a subsequent pulse, only after a previous \textbf{1} $\rightarrow$ \textbf{2} conversion, or even during the same (long) voltage pulse (Fig.~\ref{fig:schalten} bottom, see also SI Fig. S1). In Fig.~\hyperref[fig:schalten]{\ref*{fig:schalten}(c)}, it can be seen that even four molecules are changed from \textbf{2} to \textbf{1} during one $2~\mathrm{V}$ pulse. Several \textbf{1} $\rightarrow$ \textbf{2} events were performed directly before the measurement of this sequence shown in Fig.~\ref{fig:schalten} top. Back converted molecules are often located further away from the position of the tip apex and, therefore, in the close packed layer it is difficult to control which molecules of \textbf{2} are converted. The precision of the manipulation could be improved by using pulses of lower bias voltage (e.g. $1.5~\mathrm{V}$) but the conversion rate is correspondingly drastically reduced.

\section{Discussion}

Very similar cases with two different appearances of the same molecule in STM were observed before, but could not be explained adequately: Two appearances of \textit{meso}-tetra\-mesityl\-porphyrin on Cu(100), which also clearly differ from the metalated molecule, were reported by Écija \textit{et al.}~\cite{Ecija2008} Thereby, the number of ``bright'' molecules compared to ``dark'' molecules was found to decrease with scanning time. The authors tried to explain this observation with two different molecular conformations of the porphyrin, whereby a not specified higher energy conformation is irreversibly switched into a conformation with lower energy by tip influence. Furthermore, they propose that the second conformation might be due to intermolecular interactions. Indeed, switching of the porphyrin deformation was observed before for individual molecules at low coverage.\cite{Qiu2004, Iancu2006} Nevertheless, for the system investigated here, the possibility of two different molecular conformations as reason for the appearance of \textbf{1} and \textbf{2}, \textit{i.e.} different angles $\alpha\_{ph}$ and $\beta\_{pyr}$, can be ruled out by our highly resolved STM results. As mentioned in the previous section, an identical rotation of the phenyl groups at low bias voltage and the existence of a saddle-shaped deformation of the macrocycle at higher voltages is observed for both species. 
Bai \textit{et al.}~\cite{Bai2008}\ reported for the free-base phthalocyanine (H$\_2$Pc) molecule on Ag(111) two apparently different heights, whereby both kinds could be metalated by deposition of metal atoms. For the planar H$\_2$Pc, this can clearly not be explained by different molecular conformations and, therefore, a different molecule-substrate interaction was suggested as possible reason. Such a situation can be ruled out for H$\_2$THPP because the appearance of two different electronic structures for the same evaporated molecule is clearly not related to the adsorption specifically on Au(111), e.g. on different adsorption positions. We refer here to our investigation of H$\_2$THPP on Ag(110)\cite{THPPstruc}, where this effect is also observed. Furthermore, a conductance switch based on tautomerization of the central H atoms in naphthalocyanine\cite{Liljeroth2007} and H$\_2$TPP\cite{Auwaerter2012} was demonstrated before. Tautomerization is also unlikely to explain the significant change in the electronic structure between \textbf{1} and \textbf{2} because for these tautomers the measured different conductance originates from the rotation of a molecular orbital with lower symmetry than the molecule, when at low temperature and even relatively low bias voltage the tip is positioned over a pyrrole ring of H$\_2$TPP or a leg of H$\_2$Nc~\cite{Fu2009}. In the work of Auwärter \textit{et al.}~\cite{Auwaerter2012}, additionally, the subsequent removing of single H atoms is shown, which changed the molecular appearance in STM at $-0.2~\mathrm{V}$. Similarly, molecules with a single protrusion, for example the one marked with a white rectangle in Fig.~\ref{fig:schalten}, are very rarely observed and might be mono-dehydrogenated molecules or due to an asymmetric deformation\cite{Muellegger2011} of H$\_2$THPP.

\begin{figure}[htb]
\centering
\includegraphics[width=0.95\textwidth]{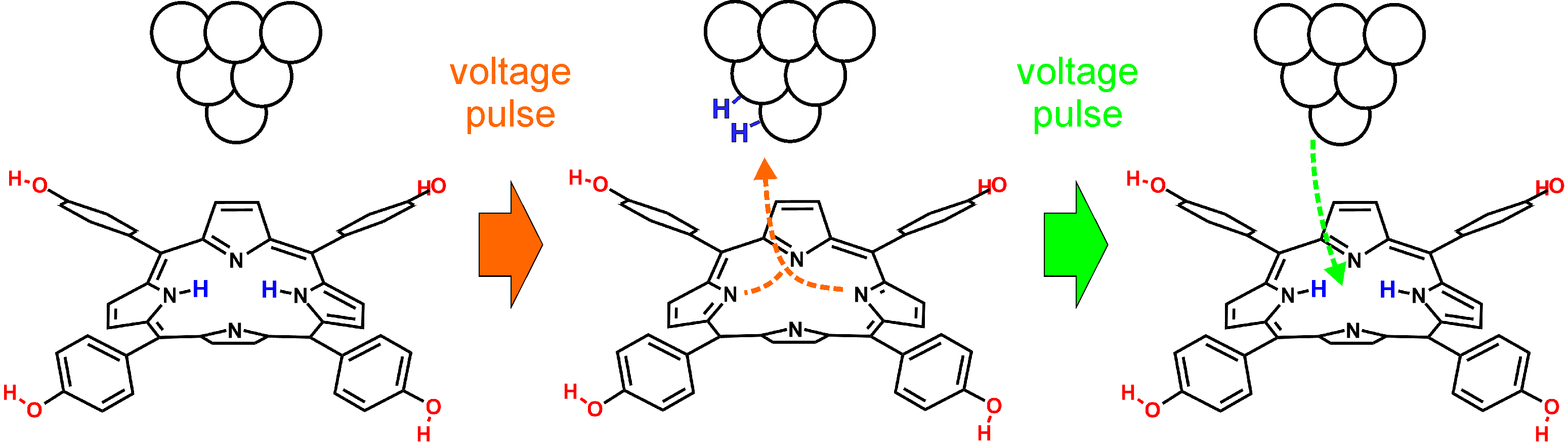}
\caption{Scheme of the proposed transfer of 2H from H$\_2$TPP to the tip apex and from the tip back to the dehydrogenated molecule by voltage pulses (\textbf{1} $\rightarrow$ \textbf{2} $\rightarrow$ \textbf{1}).}
\label{fig:skizze}
\end{figure}

Our conclusion is that during the manipulation with the STM tip at relatively high bias voltages, the central hydrogen atoms are transferred from the molecule to the tip apex~\cite{Liu2013}, but can also be removed from the tip and again incorporated into the molecule with another voltage pulse. This process is schematically shown in Fig.~\ref{fig:skizze}. The less controllable and more scattered back-conversion is then probably due a diffusion of previously transferred H atoms on the tip apex which leads to a larger impact area for subsequent pulse-induced desorption. After \textbf{1} $\rightarrow$ \textbf{2} conversion events sometimes imaging becomes fuzzy, which indicates the unstable adsorption of atoms at the tip apex. In a similar way, transfer of a Cl ligand between iron-tetraphenylporphyrin molecules using a STM tip was reported recently.\cite{Gopakumar2012a}
From our observations, we propose therefore that the molecular species found on the surface are H$\_2$THPP and the dehydrogenated molecule (THPP).
Chemically changing the molecule is found to be possible at both polarities of the bias voltage and with the probability increasing with the current value (see SI Fig. S1).
Thus, the N-H bond cleavage is likely induced by multiple-vibrational excitation due to inelastic tunneling of electrons from the tip or from the Au(111) surface into molecular orbitals.\cite{Avouris1996} The dissociated H atoms then move up and adsorb on the tip apex. The formation of a H$\_2$ molecule directly after dissociation of both H atoms would be energetically favorable and could explain why in our case during voltage pulses an interconversion between two species and not three (including singly dehydrogenated molecules) is observed.
Another explanation is the reversible dehydrogenation not of H$\_2$THPP, but of its doubly protonated form H$\_4$THPP, \textit{i.e.} with two additional hydrogen atoms bound to the pyrrolic nitrogen atoms, which then would correspond to \textbf{1}. Several porphyrins, especially those possessing sulfonate\cite{Friesen2011} or carboxyl\cite{Yoshimoto2008} groups, are found to exist in a stable protonated form after deposition from acidic solution. If this would be the case here, H$\_4$THPP $\rightarrow$ H$\_2$THPP $\rightarrow$ THPP conversion by voltage pulses should be possible. Nonetheless, as discussed, only two species with a different electronic structure near $E\_F$ could be found with STS.
The bond dissociation energy for H$\_4$THPP $\rightarrow$ H$\_2$THPP + H$\_2$ was calculated to be 211~kJ/mol (2.19~eV), which is circa the half compared to 407~kJ/mol (4.22~eV) for H$\_2$THPP $\rightarrow$ THPP + H$\_2$ (in vacuum and in absence of an external electric field). The former dissociation energy is relatively low and should lead to the  dehydrogenation of all H$\_4$THPP molecules at annealing temperatures of e.g. 250$\,^{\circ}$C, so that only one kind of molecular appearance (\textbf{2}) would be left, which however was not observed.


\begin{figure}[htb]
\centering
\includegraphics[width=0.7\textwidth]{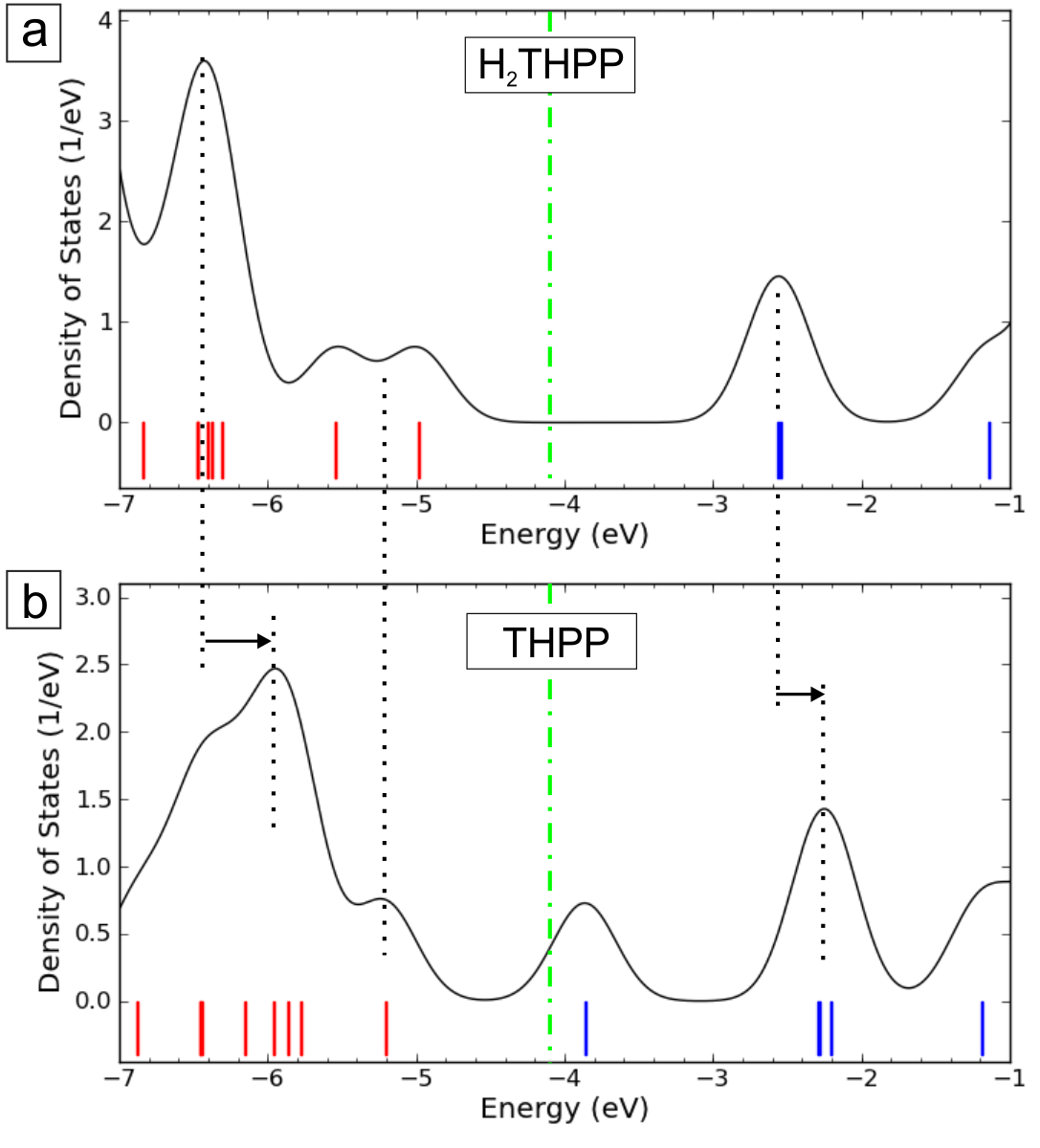}
\caption{Densities of states of H$\_2$THPP (top) and the dehydrogenated form THPP (bottom) calculated with B3LYP. Red marks occupied, blue empty molecular orbitals. The states were convoluted with a 0.3~eV wide Gaussian function to simulate the experimental broadening.}
\label{fig:DFT_deprot}
\end{figure}

To explain the shifts of the molecular orbitals between the two slightly different chemical structures of the porphyrin molecule and to corroborate our explanation, the experimental spectra are compared to DFT calculations. If the effect is indeed mostly independent of the specific substrate as observed, although the interaction of the nitrogen with surface atoms could promote a dehydrogenation, respective differences in the electronic structure should be seen also for gas-phase calculations of the molecule. Results of those are shown in Fig.~\ref{fig:DFT_deprot}. The shape of the calculated densities of states of H$\_2$THPP (top) and THPP (bottom image) is in relatively good agreement with the measured STS feature of \textbf{1} (red curve in Fig.~\ref{fig:spannungsbilder}) and \textbf{2} (green curve). If H$\_2$THPP is dehydrogenated (to THPP), the former HOMO is depopulated and all states relax to different energies. Especially, a shift of the former LUMO can be seen and the large peak from a convolution of HOMO$-$2 and subjacent states is broadened and also shifted in its energetic position similar to the measured STS of \textbf{2} (shifts are indicated by arrows in Fig.~\ref{fig:DFT_deprot}). A similar shift of the LUMO is also reported between the free-base and a metalated variant of tetra\-phenyl\-porphyrin~\cite{Comanici2008} as well as tetra\-pyridyl\-porphyrin~\cite{Zotti2007}. 
The Fermi energy of the molecule at the interface to Au(111) is determined by the work function of the substrate and the interface dipole. The Fermi energy approximated by comparison with STS ($U=0~\mathrm{V}$) is indicated by the green line in Fig.~\ref{fig:DFT_deprot}. For THPP the former HOMO, now at an energy of $-3.86~\mathrm{V}$, should be very close to $E\_F$ for the system THPP on Au(111). Nevertheless, a peak for this state could not be measured with STS for \textbf{2} (green curve in Fig.~\ref{fig:spannungsbilder}). A reason for this might be the lower sensitivity of the method at in the low bias region.
Also, when comparing experimental data with DFT, it should be noted that while GGA-DFT generally underestimates HOMO-LUMO gaps, the hybrid-DFT functional B3LYP (25\% exact exchange) gives a value too large by 0.6~eV, when compared with $(\mathbf{d}I/\mathbf{d}U)$ of H$\_2$THPP on Au(111). With smaller angles $\alpha\_{ph}$ and corresponding larger $\beta\_{pyr}$, that is going from planar to increasing saddle-shape deformation, the HOMO was found to shift toward $E\_F$, which enlarges the HOMO - HOMO$-$1 separation and decreases the HOMO-LUMO gap by around 0.4~eV for $\alpha\_{ph} = 30^{\circ}$ compared to $67^{\circ}$. Otherwise, different geometries cause only negligible changes in the energetic position of the other molecular orbitals near $E\_F$.

Summarizing, our DFT results indicate by comparison with STS that \textbf{1} and \textbf{2} corresponds to H$\_2$THPP and THPP, respectively, and therewith the underlying process, which explains the experimentally observed change in the electronic structure, is due to a reversible transfer of hydrogen between molecule and STM tip.

\section{Conclusions}

Two stable forms of 5,10,15,20-tetra(\textit{p}-hydroxyphenyl)\-porphyrin adsorbed on Au(111) could be observed with STM. They can be assigned to the free-base H$\_2$THPP and the dehydrogenated form THPP. Our tunneling spectroscopy data reveals significant changes of the energetic positions of the molecular orbitals near the Fermi energy which gives experimental insight in the relaxation process of all molecular states after dehydrogenation. Because the chemical structure of individual porphyrin molecules in a highly ordered array can be changed by hydrogen transfer to and from a STM tip at 2~V and non-manipulative read-out of the different conductance is possible at e.g. 1.2~V, this presents the possibility for an application as nano-scale memory device at room temperature based on dehydrogenation and hydrogenation of a molecule. To improve the control of addressing only one single molecule, the aim of future work would be to increase the distance between the H$\_2$THPP molecules in the array, for example by coadsorption with other molecules which act as spacer.

\section*{Acknowledgements}

This work has been financially supported by the Deutsche Forschungsgemeinschaft (DFG) through the Research Unit FOR 1154 and the Fonds der Chemischen Industrie (FCI). Computational resources were provided by the ``Chemnitzer Hochleistungs-Linux-Cluster'' (CHiC) at the Technische Universität Chemnitz.

\section*{Supporting information available}

DFT optimized molecular conformations; $I-t$ and $I-U$ spectra of inter-conversion events; Single adsorbed H$\_2$THPP molecules on top of the first layer; DOS recovery from STS and energies of the molecular orbitals; Materials, synthesis and characterization of H$\_2$THPP

\bibliographystyle{model1a-num-names}
\bibliography{THPP_spec.bbl}

\end{document}